\documentclass[twoside,twocolumn]{article}

\usepackage{blindtext} 

\usepackage[sc]{mathpazo} 
\usepackage[T1]{fontenc} 
\usepackage{microtype} 

\usepackage[english]{babel} 

\usepackage[hmarginratio=1:1,top=32mm,columnsep=20pt]{geometry} 
\usepackage[hang, small,labelfont=bf,up,textfont=it,up]{caption} 
\usepackage{booktabs} 

\usepackage{lettrine} 

\usepackage{enumitem} 
\setlist[itemize]{noitemsep} 

\usepackage{abstract} 

\usepackage{titlesec} 
\renewcommand\thesection{\Roman{section}} 
\renewcommand\thesubsection{\roman{subsection}} 
\titleformat{\section}[block]{\large\scshape\centering}{\thesection.}{1em}{} 
\titleformat{\subsection}[block]{\large}{\thesubsection.}{1em}{} 

\usepackage{fancyhdr} 
\usepackage{titling} 

\usepackage{hyperref} 

\usepackage{graphicx,amssymb,mathrsfs,amsmath,pifont,amscd,latexsym,color,fancyhdr,CJK,url,hyperref,multirow, subfigure, colortbl,tabularx,threeparttable, booktabs,balance,natbib}
\usepackage{flushend}

\pretitle{\begin{center}\Huge\bfseries} 
\posttitle{\end{center}} 
\title{PRADA Applicability in Industrial Practice} 
\author{%
\textsc{Xuan Lu}
\\[1ex] 
\normalsize Peking University \\ 
\normalsize \href{mailto:luxuan@pku.edu.cn}{luxuan@pku.edu.cn} 
}
\date{\today} 
\renewcommand{\maketitlehookd}{%
\begin{abstract}
The proliferation of Android devices brings the fragmentation problem. Selecting and prioritizing major device models are critical for mobile app developers to select testbeds and optimize various issues such as quality-assurance, release planning, revenues, and so on. PRADA, an approach proposed in our previous work, was designed to help \textbf{pr}ioritizing \textbf{A}ndroid \textbf{d}evices for \textbf{a}pps. In this paper, we conduct a survey in two IT companies to study the applicability of PRADA. Results from the survey show that PRADA has attracted the interests of industrial Android developers.
\end{abstract}
}

\begin{document}

\maketitle

\begin{table*}[htbp]
\scriptsize
  \newcommand{\tabincell}[2]{\begin{tabular}{@{}#1@{}}#2\end{tabular}}
  \centering
  \caption{Survey Questionnaire Filled in by the Companies}
  \begin{threeparttable}
    \begin{tabularx}
      {\textwidth}{X r}
       \toprule[3pt]
      \bfseries \textbf{Question} & \bfseries \textbf{Answer} \\
     \hline
   
     \textbf{Background of the  company} & \\
     \hline
      1. How many years of experiences do you have in Android development? & \textbf{Open}\\
      2. How many apps do you develop? & \textbf{Open}\\
      \hline
     \textbf{Experiences with Android fragmentation} & \\ 
     \hline
     3. Is device fragmentation a problem in your app development? & \textbf{Yes$\mid$No}\\
     4. If ``\textit{Yes}'' to 3, on what level of impacts do you rate the fragmentation problem? Additionally, can you provide any evidence? 
     \\
     5. If ``\textit{Yes}'' to 3, on what aspects/stages do you think that fragmentation can affect app development? & \textbf{Open}\\
     6.  If ``\textit{Yes}'' to 3, how do you allocate resources for dealing with prioritizing major device models in your app development? & \textbf{Open}\\

     \hline
     \textbf{Solutions of addressing Android fragmentation} &\\
     \hline  
		7. How do you identify the major device models? & \textbf{Open}\\
     8. How many device models do you take into account in your app development? & \textbf{Open}
\\
     9. Are your selected device models accurate? (\textbf{1. Yes, always; 2. Sometimes not very accurate; 3. No, not very satisfying} ) & \textbf{1$\mid$2$\mid$3}\\
     \hline
   \textbf{Feedback when having the 25-minute tutorial and demonstration of PRADA} &\\  
   \hline
   10. Do you think that techniques such as PRADA are reasonable and useful for app development? & \textbf{Yes$\mid$No}
   \\
   11. If ``\textit{Yes}'' to 10, on what aspects/stages can PRADA benefit your app development? & \textbf{Open}\\
   12. If ``\textit{Yes}'' to 10, would you like to try PRADA in your actual development?& \textbf{Open}\\
   13. If ``\textit{Not}'' to 10, why do you think that PRADA is not reasonable? & \textbf{Open}\\
   14. Do you have any suggestions on improving PRADA? & \textbf{Open}\\
       \bottomrule[3pt]
     
    \end{tabularx}
  \end{threeparttable}
  \label{feedback}
\end{table*}

\section{Introduction}
Based on our previous work (\cite{ICSE16Lu}), we move to a new research question (\textbf{RQ 4}), i.e., {\textit{would actual Android developers consider leveraging PRADA for prioritizing device models in practice?} In order to answer this question, we make a qualitative study with a semi-structured survey by interviewing the project managers and programmers from two software companies. We design some questions and analyze the practical applicability of PRADA in their actual development activities. To make our study representative, the chosen two companies have different businesses and target user groups. 

The first company is the Xinmeihutong Inc\footnote{\url{http://xinmei365.com/index_en.html}} (abbreviated as Xinmei in the rest of this article). Xinmei develops several apps including input method tools and games aiming at users worldwide. For example, one app from Xinmei, called Kika emoji keyboards, is very popular on Google Play\footnote{\url{https://play.google.com/store/apps/details?id=com.qisiemoji.inputmethod}}, with more than 10 million users worldwide and more than 50 million downloads as of June 2016. The second company is the Chelaile\footnote{\url{http://www.chelaile.net.cn/}}, which develops transportation and navigation apps mainly serving Chinese users. The active users of Chelaile has reached up to about 0.8 million.

\section{Survey Organization}
Our survey study is semi-structured. We aim to pilot the interviewees to report their requirements and experiences in dealing with device-model fragmentation. To this end, we  appreciate quite open answers and the interviewees are not displayed too many predefined options. We design a questionnaire that consists of some questions. We begin with investigating the background of the companies under study such as the experiences on Android development. Then we investigate the evidence of facing the Android-fragmentation problem in their actual development with questions such as \textit {``Is device fragmentation a problem in your app development?'';``On what level of impacts do you rate the fragmentation problem
and can you provide some evidence?''}, and so on. 

When receiving the confirmation of necessity of device-model prioritization out of fragmented Android devices, we then explore how they address such a problem, with questions such as  \textit{``How do you identify the major device models for you apps?''; ``How many device models do you need"; Are the selected device models accurate?''}, and so on. 

After getting their answers, we make a 25-minute tutorial on introducing the idea of PRADA (as presented in preceding sections). We then demonstrate the performance of PRADA on some typical apps from the same category of the apps that are developed by the company being interviewed. The interviewees are asked to assess the applicability of PRADA and whether they would like to consider applying PRADA in their actual development activities. In particular, they are free to make any comments or suggestions of PRADA.

\section{Study Results}

We next report the survey-study results from the two companies, respectively.\\

\noindent $\bullet$ \textbf{\textsc{Interview 1}: Project Manager @ Kika Tech}.

\noindent The project manager reported that Xinmei was founded in 2011 and most of their apps are designed for only the Android platform. In response to our first question, he confirmed the device fragmentation problem as a significant issue in developing Android apps: ``\textit{Our company develops various themes/skins of the Kika keyboard app and some game apps. We find that there are hundreds of Android devices, whose hardware specifications are of high diversity. In particular, the size estate and graphical definition level of device models play important roles of game apps. We need to make our apps perform correctly on every single device model.}" 

The project manager reported that device prioritization is important to their revenues. He explained  \textit{``We have urgent needs to know which device models are important to our apps. Our apps are all free, and our revenues mainly come from the in-app ads. We need to know which users are more likely to click the ads, and thus negotiate with ads network providers to design and deliver ``customized" ads and attract user clicks."} The project manager also confirmed that prioritizing the major device models is indeed a quite time-consuming task when developing new apps or releasing new app versions: \textit{ ``Certainly, we can collect the device model information in our apps, and can know which device models are major ones. However, such information can be obtained only when the new apps have been released after a while. At the very beginning of publishing these new apps, we do not have such knowledge at all. We usually develop 2-4 new game apps every year. We would like to ``predict'' who will use our apps and deliver customized ads in advance.}'' The project manager said that they used to rely on the popular device models from famous manufacturers, such as Samsung and LG, as they assumed that these device models can have higher user coverage. ``\textit{However, from the user feedback after a while, these popular device models are usually high-end ones, but are not always representative ones. For example, some game apps are quite popular on some relatively lower-end device models. In practice, we sometimes rely on the developers' personal experiences and choices. However, our developers have to spend a lot of time to decide which device models to be purchased for testing purposes. Although I can approve necessary budgets on buying 5-10 device models, I would like to have better Return-of-Investment.} "

After getting the basic knowledge of PRADA as well as its performance on various game apps, the project manager showed very strong interest: ``\textit{I know Wandoujia and its monitoring features. It is impressive that you can utilize the usage data for prioritizing device models. It can help us more accurately locate and predict the major device models that can indicate the potential users of our apps.}" The project manager confirmed that PRADA is reasonable and useful, \textit{``I think that I can understand your PRADA approach well and it makes sense to me. PRADA essentially leverages the existing knowledge of other similar or competing apps and tells the shared commonality of them. It is a cool idea, as you have the unique operational data from actual users.''} The project manager also confirmed that PRADA is quite useful in the Android development practice. When he was told that PRADA shall be deployed as an analytic service (\cite{Liu2009Discovering, ICSW15Ma}), he was extremely enthusiastic to adopt PRADA in his company: ``\textit{Yes, I am sure that I would like to try PRADA in our actual practice. I can check the major device models via PRADA periodically, e.g., monthly or quarterly. I think that PRADA can reduce the risk of useless investment on buying device models, and can save our efforts and improve targeting potential users.}''

The project manager also positively pointed out two improvements for PRADA. ``\textit{I think that PRADA essentially depends on the operational data of the existing device models, right? For the newly released device models where you do not have such data, I am not sure whether PRADA can help a lot. You may consider collecting and analyzing the operational data more frequently to provide timely results.}'' The second suggestion is to help prioritize those device models where bugs can happen:  ``\textit{In the past we check the user reviews on Google Play, where the textual reviews can be categorized by device models. However, Google Play closed such feature from 2014.}" The project manager suggested that the Wandoujia management app record the crash traces at system level, or support device-specific user review and bug categorization. He quoted \textit{``I believe that leveraging such data can enlarge the applicability of PRADA.''}

The project manager discussed the current ``localization" of PRADA, since our current evaluation and demonstration rely on Wandoujia, as the major device models can be different in different countries or areas. ``\textit{However, it may not be a big issue. If other app stores or third-party apps can have the operational profile data like yours, I think that PRADA can still work. At least, currently PRADA can help us a lot in China, the biggest market all over the world.}''\\
 
\noindent $\bullet$ \textbf{\textsc{Interview 2}:  Project Director @ Chelaile}.

\noindent Compared to Xinmei, which develops various apps that are downloaded worldwide, Chelaile is a more focused company that develops  navigation apps for only mainland China. Chelaile was founded in 2013, and has supported both iOS and Android platform since then. Similar to the project manager at Xinmei, the project director at Chelaile also confirmed that Android device fragmentation is a serious problem in their development activities (``\textit{\textbf{absolutely yes}}''). The project director complained that identifying and selecting major device models are a non-trivial task, and quoted ``\textit{When releasing every new version of our apps, we need to make sure that our apps can work properly on every possible device model. It would be very important to know the device models where one of our apps can have bugs before the app is released. Additionally, we also care about the device models where our users come from. However, it took us a lot of time, as there are numerous device models on market.}" Similar to Xinmei, the project director's company relied on the market-share report, the developers' personal experiences, and the device-model information collected in their historical app versions, to select the device models for testing. However, as a new startup company, the project director's company has very limited budget to buy so many device models. ``\textit{We choose to use the cloud-side VM to test our apps on device models. It is a much economically practical and affordable solution}.'' However, the project director also confirmed that such a way of device-model prioritization can save only the expense on buying physical device models; it cannot promise high coverage of bugs and users, and the efforts of prioritizing device models is not alleviated at all. 

Although the demonstration of PRADA did not relate to predicting bugs, the project director was interested in PRADA and confirmed its validity in their actual development activities. \textit{``Our previous experiences showed that our apps can have similar problematic issues with other (navigation) apps on some specific device models. The PRADA approach sounds reasonable since you have useful data from millions of users who use similar apps. The analytics can be statistically meaningful and can be a good indicator for guiding us to prioritize the major device models. If PRADA can incorporate the bug reports of similar apps, maybe some possible bugs of our apps can be predicted to exist on specific device models. Hence, the prioritization is still expected."}  The project director was also willing to try PRADA in their practice. \textit{``The browsing-time case of PRADA can help us prioritize the major device models that stand for more users.''} Compared to the project manager's company, which tries to explore PRADA for in-app ads, the project director regarded that PRADA could bring additional benefits for a release-planning strategy. ``\textit{We consider that the choice of device model can somehow reflect the possible (economic) background of a user. Users holding lower-end device models would like to save data plan. For these users, we can provide additional features such as offline map and special data plan.
If we can predict how many users a device model can own, we can negotiate with its manufacturers and provide a special app version, or even make our apps preloaded. Indeed, PRADA may be also useful to optimize our investment and resource allocation of release planning.}''

Finally, the project director considered that improving PRADA's effectiveness can be further explored in terms of incorporating finer-grained data, e.g., the OS versions used on device models.\\

\noindent $\bullet$ \textbf{Findings}. Based on the feedback from our interviews, we can conclude that PRADA has attracted the interests of industrial Android developers. Both the two companies confirmed the applicability of PRADA as an analytics service and had the interest to apply PRADA in their development activities. The feedback also indicated that the fragmentation of Android devices can affect various aspects including not only the incompatibility or bugs, but also the revenues and the release of new features. Collecting and leveraging the operational usage data with PRADA can have a wide spectrum to alleviate the issues caused by fragmentation. Additionally, the feedback also provided us some potential improvements in future work~(\cite{IMC15Li, TSE17Liu, TOIS17Liu}). For example, the PRADA approach can be designed and provided as web service (\cite{SOSE13Ma, Splash12Wang, Chinaf14Liu, ICWS15Liu}) to Android developers.

\bibliographystyle{ACM-Reference-Format-Journals}
\bibliography{temp}

%
%


\end{document}